\documentclass[reprint, 
 amsmath,amssymb,
 aps, physrev,
]{revtex4-2}

\usepackage{graphicx}% Include figure files
\usepackage{dcolumn}% Align table columns on decimal point
\usepackage{bm}% bold math
\usepackage[colorlinks=true,
            linkcolor=blue,
            citecolor=blue,
            urlcolor=blue,
	    breaklinks=true]{hyperref}% add hypertext capabilities
\usepackage{tikz-feynman}
\usepackage{url}

\def\bnon{\begin{equation*}}
\def\enon{\end{equation*}}
\def\bnu{\begin{equation}}
\def\enu{\end{equation}}
\def\bqa{\begin{eqnarray*}}
\def\eqa{\end{eqnarray*}}
\def\blsd{\begin{enumerate}}
\def\elsd{\end{enumerate}}
\def\blst{\begin{itemize}}
\def\elst{\end{itemize}}
\def\eps{\varepsilon}
\def\ide{\mathbb{I}}

\def\eqbk{\\[3pt]}

\def\tr{\text{Tr}}
\def\neqqcd{$\mathcal{N}=0$ }
\def\neqt{$\mathcal{N}=2$ }
\def\neqo{$\mathcal{N}=1$ }
\def\neqf{$\mathcal{N}=4$ }
\def\neqa{$\mathcal{N}=1,2,4$ }
\def\neqsafe{$\mathcal{N}=1,4$ }
\newcommand{\specul}[2]{}

\newcommand{\glyuk}{\begin{minipage}{1.2cm}
\begin{tikzpicture}

        \begin{feynman}
             \vertex (o) at (0,0);
             \vertex (a) at (-0.24,-0.18);
             \vertex (b) at (0.24,-0.18);
             \vertex (d) at (0, 0.75);
             \vertex (e) at (-0.6, -0.45);
             \vertex (f) at (0.6, -0.45);
             \diagram* {
             (e) -- [fermion] (a);
             (f) -- [fermion] (b);
             (o) -- [scalar] (d);
             };
          \end{feynman}
          \filldraw[fill=gray!50, draw=black] (0,0) circle (0.3cm);
          \node (0,-0.5) {};
\end{tikzpicture}
\end{minipage}}

\newcommand{\glccg}{\begin{minipage}{1.2cm}
\begin{tikzpicture}

        \begin{feynman}
             \vertex (o) at (0,0);
             \vertex (a) at (-0.24,-0.18);
             \vertex (b) at (0.24,-0.18);
             \vertex (d) at (0, 0.75);
             \vertex (e) at (-0.6, -0.45);
             \vertex (f) at (0.6, -0.45);
             \diagram* {
             (e) -- [ghost] (a);
             (f) -- [ghost] (b);
             (o) -- [decorate,
    decoration={coil, amplitude=1.25pt, segment length=2.8pt}] (d);
             };
          \end{feynman}
          \filldraw[fill=gray!50, draw=black] (0,0) circle (0.3cm);
          \node (0,-0.5) {};
\end{tikzpicture}
\end{minipage}}

\begin{document}

\preprint{APS/123-QED}

\title{
Dimensional reduction is supersymmetric at three loops
}
\author{Mrigankamauli Chakraborty}
 \email{mrigankamauli.chakraborty@desy.de}
\author{Sven-Olaf Moch}%
 \email{sven-olaf.moch@desy.de}
\affiliation{%
II. Institute for Theoretical Physics,  Universität Hamburg \\
Luruper Chaussee 149, 22761 Hamburg, Germany
}%


\date{\today}% It is always \today, today,

\begin{abstract}
We resolve the long‑standing claim that regularisation by dimensional reduction (DR) fails to preserve supersymmetry in Super Yang–Mills (SYM) theories at three loops. 
Earlier results reported a mismatch between the Yukawa and ghost–gluon $\beta$ functions in $\mathcal{N}=2$ SYM, suggesting a breakdown of supersymmertry. 
We show that this discrepancy does not originate from DR itself but from subtleties in the treatment of the Clifford algebra. 
A corrected three‑loop calculation restores full supersymmetric behaviour, and we demonstrate that the same issue would first affect $\mathcal{N}=4$  SYM only at five loops, consistent with existing four‑loop results.
Our findings confirm that DR preserves supersymmetry for $\mathcal{N}=1, 2$ and $4$ SYM through the loop orders examined.

\end{abstract}

\maketitle

\section{\label{sec:level1}Introduction}

Supersymmetry (SUSY) is a proposed symmetry relating bosons and fermions, pairing particles of differing spin into common multiplets. 
In supersymmetric gauge theories, this symmetry imposes strong constraints on quantum corrections, often leading to improved ultraviolet behaviour and, in some cases, exact non‑renormalisation theorems, see, e.g.~\cite{Martin:1997ns}.
A central quantity for assessing these quantum properties is the $\beta$ function,
\begin{equation}\label{eq:beta-def}
    \mu\frac{d}{d\mu} a(\mu) = \beta(a) 
\, .
\end{equation}
It describes how a coupling $a(\mu)$ evolves with the renormalisation scale $\mu$, and SUSY typically requires that all couplings related by SUSY transformations run identically. 
Any mismatch in their $\beta$ functions is therefore a direct and unambiguous signal of SUSY breaking at the quantum level, either physical breaking or an artifact of the regularisation scheme.

Supersymmetric Yang–Mills (SYM) theories are based on the supersymmetric vector multiplet and are symmetric under a local non‑Abelian gauge group $SU(N)$. 
They come in several versions, depending on the number of supercharges  $\mathcal{N}$  in the SUSY algebra. 
Correspondingly, we have $\mathcal{N}=1, 2$ and $4$ vector supermultiplets. 
The \neqo SYM multiplet contains one gauge boson and one Weyl fermion; the \neqt multiplet contains one gauge boson, two Weyl fermions, and two real scalars; and the \neqf multiplet contains one gauge boson, four Weyl fermions, and six real scalars. 
The $\mathcal{N}=8$ vector multiplet contains gravitons and thus does not define an SYM theory. 
All component fields transform in the adjoint representation of the gauge group.

These increasing field contents impose increasingly rigid symmetry structures: \neqo allows for nontrivial renormalisation, \neqt already exhibits strong constraints such as the one‑loop exactness of the gauge $\beta$ function~\cite{Novikov:1983uc}, and \neqf SYM is ultraviolet finite, with  vanishing $\beta$ functions to all loop orders~\cite{Avdeev:1980bh,Grisaru:1980nk}.
Because of these properties, SYM theories have long served as testing grounds for regularisation schemes and precision checks of SUSY at the quantum level.

To compute the $\beta$ function and the other renormalisation group functions explicitly for SYM, we require a regularisation procedure that maintains manifestly both the gauge symmetry and SUSY. 
Dimensional regularisation (DREG)~\cite{tHooft:1972tcz,Bollini:1972ui} in $D=4-2\eps$ is a successful procedure for gauge theories, and a supersymmetric variant, dimensional reduction (DR), was proposed in 1979~\cite{Siegel:1979wq}, where the number of scalars $n_s$ is analytically continued to $n_s+2\eps$. Shortly thereafter, Siegel published a follow‑up paper highlighting mathematical inconsistencies in the definition of DR~\cite{Siegel:1980qs}. 
The first explicit three‑loop calculation implementing DR for the 
$\mathcal{N}=1, 2$ and $4$ SYM theories appeared in 1982~\cite{Avdeev:1981ew,Avdeev:1982np}. 
In that work, it was observed that the $\beta$ function for the Yukawa coupling 
$\beta_{ffs}$ did not match the one for the ghost–gluon coupling $\beta_{ccg}$ in any of the theories, despite SUSY-imposed constraints that these couplings should run identically:
\begin{equation}\label{eq:betas}
    \beta(a) = \beta_{ffs}(a)\left(\;\glyuk\;\right) = \beta_{ccg}(a)\left(\;\glccg\;\right)
\, .
\end{equation}
A mismatch between $\beta_{ffs}$ and $\beta_{ccg}$, i.e. a violation of eq.~\eqref{eq:betas}, signaled a breakdown of SUSY.

A few decades later, in 2009, the three‑loop calculation was repeated in~\cite{Velizhanin:2008rw} (see also~\cite{Chestnov:2019bed}). 
The updated results showed that the three‑loop Yukawa and ghost–gluon vertices behave supersymmetrically in \neqsafe SYM, but not in \neqt SYM, while consistent SUSY behaviour was confirmed for all other three‑point vertices, with the sole exception of the \neqt Yukawa coupling as summarized in tab.~\ref{tab:wrongres}. 
In 2011, the calculations in~\cite{Velizhanin:2010vw} further established that the four‑loop $\beta$ functions for all vertices in \neqf SYM vanish, confirming fully consistent supersymmetric behaviour.
\begin{table*}
\caption{\label{tab:wrongres} Results for the gauge and Yukawa vertices of \cite{Velizhanin:2008rw}. 
    Here, $ccg$ refers to the ghost($c$)-gluon($g$) vertex and $ffs$ refers to the fermion($f$)-scalar($s$) Yukawa vertex, see eq.~\eqref{eq:betas}.
    $C_A$ and $d_{44}$ denote the adjoint quadratic and quartic Casimir invariants of the $SU(N)$ gauge group~\cite{vanRitbergen:1998pn}.
    The result for the \neqt Yukawa $\beta_{ffs}$ function violates SUSY, see~\cite{Novikov:1983uc}.}
\begin{center}
\begin{ruledtabular}
\begin{tabular}{ c c c }
 $\mathcal{N}=$ &  $\beta_{ccg}(a)$ & $\beta_{ffs}(a)$ \eqbk
 \hline
 1 &  $-\left(6 a^2 C_A+12 a^3 C_A^2+42 a^4 C_A^3\right)$ & $-\left(6 a^2 C_A+12 a^3 C_A^2+42 a^4 C_A^3\right)$ \eqbk 
 2 &$-4a^2C_A$ & $-4 a^2 C_A-8a^4\left( C_A^3 (4-3 \zeta_3)+d_{44}
   (12+72 \zeta_3)\right)$ \eqbk  
 4 &  $0$ & 0 \\
\end{tabular}
\end{ruledtabular}
\end{center}
\end{table*}

The violation of SUSY in the \neqt Yukawa $\beta$ function was attributed to an inconsistency of DR. 
However, this raised an unresolved question: why did DR fail only for \neqt SYM, but not for \neqsafe? 
The situation was paradoxical, the minimal \neqo theory and the maximally extended \neqf theory were protected from these inconsistencies, while the intermediate \neqt case was not. 
Moreover, no explanation was given for why all other vertices and propagators remained supersymmetrically consistent, or why the mismatch appeared only at three loops.

In the following, we provide answers to these questions. We show that the issue originates from an inconsistent handling of the Clifford algebra in the calculation procedure. 
This inconsistency leaves all three‑loop vertices and propagators unaffected except for the three‑loop \neqt Yukawa coupling, and once corrected, the \neqt Yukawa $\beta$ function becomes fully supersymmetric. 
We further demonstrate that in \neqf SYM the same inconsistency would first appear only at five loops, explaining why the four‑loop result of \cite{Velizhanin:2010vw} remained supersymmetric.

\if{1=0}

\
Studies of SUSY at the quantum level requires regularisation schemes, in particular of the perturbative approach.
Soon after proposing supersymmetric dimensional regularisation via dimensional reduction (DR) in 1979 [Siegel:1979wq], Siegel followed up with a paper discussing mathematical inconsistencies in the definition of DR [Siegel:1980qs]. The first explicit three‑loop calculation implementing DR for the N=1, N=2, and N=4 SYM theories appeared in 1983 [Avdeev:1982xy]. In that work, it was observed that the beta function for the Yukawa coupling did not match the beta function for the ghost–gluon coupling in any of the theories. Since these couplings should run identically in a supersymmetric scheme, the mismatch signalled a breakdown of SUSY.

On top of that, as predicted by the NSVZ $\beta$ function~\cite{Novikov:1983uc}, the \neqt SYM $\beta$ should only get a one-loop correction, while \neqf SYM is conformal with $\beta=0$.

Soon after his proposal of supersymmetric dimensional regularisation via dimensional reduction(DR) in 1979 \cite{Siegel:1979wq}, Siegel followed up with a paper discussing the mathematical inconsistencies in the definition of DR \cite{Siegel:1980qs}. The first explicit  3-loop calculation implementing DR for the $\mathcal{N}=1,2,4$ SYM theories was then done in 1983 \cite{Avdeev:1982xy}. There, it was observed that the $\beta$ function for the Yukawa coupling did not match the $\beta$ for the ghost-gluon coupling for either of the $\mathcal{N}=1,2,4$ theories. This is a breakdown of SUSY behaviour, as the couplings are expected to run identically.

About two decades later, in 2009, the 3-loop calculation was repeated in \cite{Velizhanin:2008rw},
\cite{Velizhanin:2008rw}, the results of which revealed that the 3-loop Yukawa and the ghost-gluon vertices behave supersymmetrically for $\mathcal{N}=1,4$ SYM, but not for \neqt SYM. tab.~\ref{tab:wrongres} summarizes the results from \cite{Velizhanin:2008rw} for the Yukawa and the ghost-gluon vertex. Similar SUSY consistent results were obtained in \cite{Velizhanin:2008rw} for all the other 3-point vertices in the theories, except only for the \neqt Yukawa coupling. In 2011, the calculations in \cite{Velizhanin:2010vw} established that the 4-loop $\beta$ function for all vertices in \neqf SYM also vanishes, portraying SUSY consistent behaviour.

In the following, we will be providing the answers to all these questions. It will turn out that the mistake stems from a wrongful implementation of Clifford Algebras in the computation procedure. Such an error does not affect any of the 3-loop vertices and propagators, except for the 3-loop \neqt Yukawa. Fixing that mistake gives SUSY consistent result also for the \neqt Yukawa $\beta$ function. We will also show that for $\mathcal{N}=4$ SYM such an inconsistency with the wrong Clifford algebra implementation will only show up at 5 loops, explaining why the 4-loop result in \cite{Velizhanin:2010vw} was SUSY consistent.

\section{Background}
Supersymmetric Yang-Mills (SYM) theories are based on the supersymmetric vector multiplet and are symmetric under a local non-Abelian gauge group. Based on the number of supercharges $\mathcal{N}$ in the SUSY algebra of the theory, we can have \neqa vector supermultiplets. The \neqo SYM multiplet has 1 gauge boson and one Weyl fermion, the \neqt multiplet has 1 gauge boson, 2 Weyl fermions and 2 real scalars, the \neqf multiplet has 1 gauge boson, 4 Weyl fermions and 6 real scalars. The $\mathcal{N}=8$ vector multiplet contains gravitons and hence is not considered with SYM theories. All the component fields transform under the adjoint representation of the gauge group. 

Supersymmetry imposes constraints on the theories. In all of \neqa SYM theories, the Yukawa coupling should run identically to the gauge coupling, e.g.
\begin{equation}\label{eq:betas}
    \beta_{ffs}\left(\;\glyuk\;\right) = \beta_{ccg}\left(\;\glccg\;\right)
\end{equation}
between the Yukawa and ghost-gluon vertex, 
i.e.
\begin{equation}\label{eq:beta-ward-id}
    \beta(a) = \beta_{ffs} = \beta_{ccg} 
\end{equation}
On top of that, as predicted by the NSVZ $\beta$ function~\cite{Novikov:1983uc}, the \neqt SYM $\beta$ should only get a one-loop correction, while \neqf SYM is conformal with $\beta=0$.

To compute the $\beta$ function and the other RG functions explicitly for SYM, we require a regularisation procedure that maintains manifestly both the gauge and super-symmetry. Dimensional regularisation (DREG) is a successful procedure for gauge theories, and a prescription for supersymmteric DREG via dimensional reduction (DR) wherein the number of scalars $n_s$ is analytically continued to $n_s+2\eps$ was laid down in \cite{Siegel:1979wq}. 

Explicit higher loop calculations implementing DR that exist in the literature (tab.~\ref{tab:wrongres}), have shown that for \neqsafe SYM the $\beta$ function follows eq.~\eqref{eq:betas} as expected by SUSY. Notice that $\beta_{ccg}$ for \neqt SYM receives only a one-loop correction, consistent with \neqt supersymmetry. We also see the expected conformal symmetry ($\beta = 0$) for \neqf SYM. But for \neqt SYM, the Yukawa $\beta$ violates SUSY constraints at 3-loops. This is the problem that we will be fixing in this work.

\fi

\section{The generalised Lagrangian}

Under DR, the continuation from $D=4$ to $D=4-2\eps$ dimensions is implemented by analytically continuing the number of scalar fields as $n_s\rightarrow n_s+2\eps$. 
Accordingly, the starting point must be a Lagrangian formulated for an arbitrary number of scalars $n_s$. 
We can therefore write the following generalised Lagrangian for the SYM theories:
\begin{align}
\mathcal{L}^{\text{gen}}_{\text{SYM}} =& \text{Tr}\bigg[-\frac{1}{4}F^{\mu\nu}F_{\mu\nu} + i\bar\lambda_A\bar\sigma^\mu D_\mu\lambda^A\nonumber + \frac{1}{2\xi}(\partial^\mu A_\mu)^2\\ 
&+\partial^\mu \bar{c} D_\mu c+\frac{1}{2}D_\mu\phi^ID^\mu\phi_I+\frac{1}{4}g[\phi^I,\phi^J][\phi_I,\phi_J] \nonumber \\
    &+g\,\phi_IZ^I_{AB}\lambda^A\lambda^B+g\,\phi_I (\bar Z^I)^{AB}\bar\lambda_A\bar\lambda_B
    \bigg]
\end{align}
where the trace is over the colour indices in the adjoint representation of $SU(N)$ under which the fields transform, with $F_{\mu\nu}^a$ the gauge field strength tensor, $\lambda^a_A, \bar\lambda^{A}_a$ are the Weyl spinors, and $\phi^I$ are real scalars. Note that upper/lower does not matter for the colour index $a$ as it is a real representation. The $\{I,J\}$ and $\{A,B\}$ are flavour indices for the fermions and scalars respectively, with $n_s = \delta^I_I$ and $n_f=\delta^A_A$. We express the perturbative expansion in terms of the coupling $a=(g/4\pi)^2$.
Results of all the \neqa SYM theories emerge from this one Lagrangian at specific values of $(n_s,n_f)$.
A similar such generalised Lagrangian for the Gross-Neveu-Yukawa model was used in \cite{emgsusy}, where the number of scalars and fermions $(n_s,n_f)$ could be varied.

As there are no evanescent interaction terms in the SYM theories, so must be in $\mathcal{L}^{\text{gen}}_{\text{SYM}}$. 
It was shown in \cite{emgsusy}, that the flavour elements $Z^I,\bar Z^J$ then must satisfy the following relation:
\begin{equation}\label{eq:flavsigma}
    Z^I\bar Z^J+Z^J\bar Z^I = 2\delta^{IJ}\ide_Z
    \;,
\end{equation}
where the identity $\ide_Z$ is the same size as the $Z^I$'s. This relation completely fixes the flavour factors of the diagrams in terms of $n_s,n_f$, as the Feynman diagrams only depend on the trace of alternating even products of the $Z^I,\bar Z^J$. 
Here, we briefly summarise the trace identities (see~\cite{emgsusy} for details) for positive integer values of $n_s$, before we discuss the extension to the regularised case. 
For even  $n_s>0$, we can construct 
\begin{equation}\label{eq:gamma}
    \Gamma^I = \begin{pmatrix}
    0 & Z^I\\\bar Z^I&0
\end{pmatrix}
\, .
\end{equation}
The $\Gamma^I$'s then satisfy the Clifford algebra:
\begin{equation}\label{eq:flavclifford}
    \Gamma^I\Gamma^J+\Gamma^J\Gamma^I = 2\delta^{IJ}\ide_\Gamma\;.
\end{equation}
and we can define the analogue of $\gamma_5$ for even $n_s>0$ as:
\begin{equation}\label{eq:gamma5}
\Gamma_{(n_s+1)} =
     \dfrac{i^{\frac{n_s}{2}}}{n_s!}\epsilon_{I_1\ldots I_{n_s}}\Gamma^{I_1}\ldots\Gamma^{I_{n_s}} 
\, ,
\end{equation}
where the overall phase arises from the Euclidean metric in the Clifford algebra in eq.~\eqref{eq:flavclifford}.
The even alternating traces of the  $Z^I,\bar Z^J$ can then be obtained as
\begin{equation}\label{eq:tracesig}
\begin{split}
    \tr[Z^{I_1}\ldots \bar Z^{I_{2n}}]= \frac{1}{2}\tr[\Gamma^{I_1}\ldots\Gamma^{I_{2n}}(1+\Gamma_{(n_s+1)})]
\, ,
\\
     \tr[\bar Z^{I_1}\ldots Z^{I_{2n}}]= \frac{1}{2}\tr[\Gamma^{I_1}\ldots\Gamma^{I_{2n}}(1-\Gamma_{(n_s+1)})]
\, .
\end{split}
\end{equation}
The part of the trace dependent on $\Gamma_{(n_s+1)}$ denotes the \emph{chiral} part, and the remaining is the \emph{achiral} one. 
Traces with odd number of flavour elements are not relevant for the theory as there are no diagrams that generate them (see, again~\cite{emgsusy} for details). 
We list here a few specific cases relevant for the three-loop calculation:
\begin{equation}\label{eq:neq2}
    \tr[Z^I\bar Z^J] = 
    \begin{cases}
        \tr[\ide_Z](\delta^{IJ}+i\epsilon^{IJ})&\text{for $n_s = 2$} 
    \, ,\\
        \tr[\ide_Z]\delta^{IJ}&\text{for $n_s > 2$}
    \, ,
    \end{cases}
\end{equation}
where the chiral contribution does not ``unlock" for higher values of $n_s$, because there are too few $\Gamma$ matrices in the trace of eq.~\eqref{eq:tracesig} to produce a non-vanishing Levi-Civita tensor.
Similarly, in \neqf SYM with $n_s = 6$, there will be no flavour Levi-Civita tensors unless we have a trace of six $Z^I$'s, which does not occur before five loops.

The case of \neqo SYM ($n_s=0, n_f=1$) is different: the representation being one-dimensional over $\mathbb{R}$ implies that it is a real representation, with $Z^I=\bar Z^I$. 
Hence, eq.~\eqref{eq:flavsigma} reduces to the usual Clifford algebra in eq.~\eqref{eq:flavclifford} for $n_s=0$. Thus, there is no chiral part in the flavour traces of \neqo SYM.

\paragraph*{\textbf{Regularisation.}}
The trace relations above are for exact integer dimensions. However, under DR, the number of scalars becomes complex.
The achiral part of the traces extends straightforwardly to the regularised case. 
It is in the chiral part that we encounter the $\gamma_5$ problem.

Fortunately, at three loops, contributions from the Levi-Civita tensor are limited to simple poles; accordingly, a simplified regularisation can be employed~\cite{Davies:2019onf}:
\begin{equation}
    \Gamma_{(n_s+1)} =
     \dfrac{i^{\frac{n_s}{2}}}{n_s!}\epsilon_{I'_1\ldots I'_{n_s}}\Gamma^{I'_1}\ldots\Gamma^{I'_{n_s}} +\mathcal{O}(\eps) \, ,\,\, \text{even $n_s\geq 2$}\, .
\end{equation}
Here, the indices $I'$ denote quantities with exact integer dimensions. 
Consequently, the primary challenge in supersymmetric regularisation reduces to addressing the $\gamma_5$ problem. 
At higher loop orders, it might be useful to employ the systematic framework of Weyl consistency conditions~\cite{Poole:2019txl,Poole:2019kcm}, which has been effectively applied to the computation of the Standard Model gauge $\beta$ function at four loops~\cite{Davies:2019onf}.

For \neqo SYM, since there are no chiral components in the traces, we do not expect any fundamental obstacle in regularising the flavor factors at any loop order. 
Since we are using Weyl fermions, the traces from fermion loops involve Pauli matrices instead of Dirac matrices. In this case, the Levi-Civita tensor appears in traces involving four or more matrices:
\begin{equation}
    \begin{split}
        \tr[\sigma^{\mu}\bar \sigma^{\nu}\sigma^{\rho}\bar \sigma^{\lambda}] &= 2[(\delta^{\mu\nu}\delta^{\rho\lambda}-\ldots)-\epsilon^{\mu\nu\rho\lambda}]\\
        \tr[\bar\sigma^{\mu} \sigma^{\nu}\bar\sigma^{\rho} \sigma^{\lambda}] &= 2[(\delta^{\mu\nu}\delta^{\rho\lambda}-\ldots)+\epsilon^{\mu\nu\rho\lambda}]\;.
    \end{split}
\end{equation}
Note that the Pauli matrices above are Wick-rotated, and the simplified regularisation for the $\epsilon$-tensor is used.

\section{Restoring SUSY}

Since the computational steps in \cite{Avdeev:1982xy, Velizhanin:2008rw} are not laid out in detail, it is not possible to conclusively identify the calculational subtleties there. 
Nonetheless, we succeeded in reproducing the results shown in tab.~\ref{tab:wrongres} by excluding the Levi-Civita tensor in either the flavour traces, the Pauli traces, or both. 
In this section, we will demonstrate that, despite these subtleties with the Levi-Civita tensor, both \neqsafe SYM theories and all vertices in \neqt SYM, except for the Yukawa vertex, yield results consistent with SUSY. Subsequently, we will present the correction to the \neqt Yukawa vertex.

The Levi-Civita tensor can contribute a non-zero value to a Feynman diagram only if it is fully contracted with another Levi-Civita tensor within the diagram. Consequently, it affects only diagrams containing an even number of (open or closed) fermion chains. Up to three loops, the maximum number of such chains is two. Therefore, in this section, we will focus solely on these diagrams. 
\begin{figure}[t!]
\includegraphics[width=0.49\linewidth]{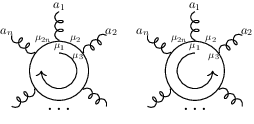}
\includegraphics[width=0.49\linewidth]{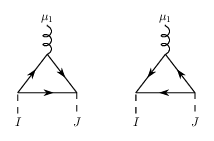}
\caption{\label{fig:epsart} Left: fermion loop connected to gluons only (arrows indicate fermion flow). Right: fermion loop with scalars.}
\end{figure}

Consider first the case of closed fermion chains (i.e., loops) that are connected solely to gluons (fig.~\ref{fig:epsart}, left). 
In this class of diagrams, the fermion flow proceeds in both clockwise and anti-clockwise directions; however, these contributions are multiplied by the same integrand. 
Therefore, the sum of these diagrams is proportional to:
\begin{align}
    \tr[F^{a_1}\ldots &F^{a_n}]\tr[\sigma_{\mu_1}\ldots\bar\sigma_{\mu_{2n}}]\nonumber\\ &+(-1)^n\tr[F^{a_n}\ldots F^{a_1}]\tr[\bar\sigma_{\mu_{2n}}\ldots\sigma_{\mu_{1}}]\nonumber\\[3pt]
    = \;\ldots \;&+\;(-1)^n\tr[(F^{a_1})^T\ldots (F^{a_n})^T]\tr[\sigma^T_{\mu_{1}}\ldots\bar\sigma^T_{\mu_{2n}}]\nonumber\\[3pt]
    = \;\ldots \;  &+\;(-1)^{4n}\tr[F^{a_1}\ldots F^{a_n}]\tr[\bar\sigma_{\mu_{1}}\ldots\sigma_{\mu_{2n}}]\nonumber\\[3pt]
    =\; (\text{\#col})&\;(\tr[\sigma_{\mu_1}\ldots\bar\sigma_{\mu_{2n}}]+\tr[\bar\sigma_{\mu_{1}}\ldots\sigma_{\mu_{2n}}])
\, ,
\end{align}
where $F^a$ is the colour matrix in the adjoint representation. We employed the invariance of traces under transpose, along with the identities $(F^{a_i})^T=-F^{a_i}$ and $\sigma_{\mu}^T=(-i\sigma_2)\bar\sigma_\mu(-i\sigma_2)$. 
Substituting eq.~\eqref{eq:tracesig} into the last line, it becomes evident that the Levi-Civita contribution cancels out. 
Therefore, in the case of a fermion loop connected solely to gluons, subtleties associated with the chiral part of the traces will have no effect.

The situation differs when scalars are involved. For instance, consider the triangle loop depicted on the right of fig.~\ref{fig:epsart} in \neqt SYM. 
Upon summing over both orientations of the fermions in the loop, the Levi-Civita contributions do not cancel:
\begin{align}
   &\tr[\sigma_{\mu_1}\ldots\bar\sigma_{\mu_{4}}]\tr[Z^I\bar Z^J]+\tr[\bar\sigma_{\mu_{1}}\ldots\sigma_{\mu_{4}}]\tr[\bar Z^IZ^J]\nonumber\\
   &\propto (\ldots - \epsilon_{\mu_1\ldots\mu_4})(\ldots + i\epsilon^{IJ})+(\ldots + \epsilon_{\mu_1\ldots\mu_4})(\ldots - i\epsilon^{IJ})
\, .
\end{align}
This illustrates a general principle: for Levi-Civita terms to survive cancellation, chiral contributions must be present in both the flavour and spinor traces of the fermion loop. 
This is because the double negatives multiply to produce a positive, preventing the chiral terms from canceling. 
It is important to note that in \neqo SYM, there are no flavour chiral terms, while in \neqf SYM, the flavour chiral contributions only become relevant at five loops. 
This accounts for why, even with an inconsistent implementation of the Levi-Civita tensor, the three-loop results in these theories remain unaffected.

In the majority of diagrams containing fermion loops, as in fig.~\ref{fig:epsart} on the right, the Levi-Civita tensors still vanish in the integrand up to three loops. 
Using the infrared rearrangement procedure \cite{Vladimirov:1979zm,Chetyrkin:1982nn} for renormalising vertices, the momentum of the external bosonic leg is set to zero. 
This effectively equalizes the two fermion momenta at that vertex, enabling the use of the relation $p^\mu p^\nu\sigma^\mu\bar\sigma^\nu = p^2$, 
which reduces two Pauli matrices within the trace. 
When this reduction occurs in a trace of four $\sigma^\mu$ matrices, the Levi-Civita contribution cancels out.

\begin{figure}[t!]
\includegraphics[width=\linewidth]{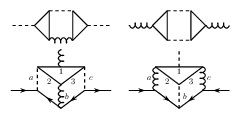}
\vspace*{-8mm}
\caption{\label{fig:epsar} 
Sample topologies of diagrams that yield a non‑vanishing Levi-Civita contribution in the integrand with gluons, fermions (solid lines, arrows indicate fermion flow), and scalars (dashed lines).
\vspace*{-1mm}}
\end{figure}
This leaves us with the four types of diagrams shown in fig.~\ref{fig:epsar}, in which the Levi-Civita tensor remains present at least within the integrand. 
For each of the diagrams in the bottom row of fig.~\ref{fig:epsar}, there are a total of $2 \times 3 \times 3 = 18$ diagrams, corresponding to the three choices of placing the external vertex on lines 1, 2, or 3, the three permutations of the bosonic lines $a, b, c$ and the two possible directions of fermion flow.

After integration, the two propagator corrections and the gluon vertex correction also vanish due to the symmetries of the integral. 
Only the Yukawa vertex diagrams (bottom right) yield a non‑vanishing Levi-Civita contribution. 
The values of the corresponding integrals are identical under permutations of $a,b,c$ once the external vertex is fixed. 
Let $I^{(n)}$ denote the value of the diagram with the external vertex on line $n$. 
Then, with the adjoint quadratic and quartic Casimir invariants $C_A$ and $d_{44}$
\begin{equation}
    \begin{split}
    I^{(1)} &= i\,(8d_{44}+\frac{2}{3}C_A^3)\frac{1}{\eps} + \mathcal{O}(\eps^0) 
    \, ,\\
    I^{(2)}= I^{(3)} &= i\,(8d_{44}-\frac{1}{3}C_A^3)(-\frac{1}{3}+\zeta_3)\frac{1}{\eps} + \mathcal{O}(\eps^0)
    \, .
    \end{split}
\end{equation}
Summing over the three permutations of $a,b,c$ gives 
\begin{align}
    3(&I^{(1)}+I^{(2)}+I^{(3)})\nonumber\\
    &=i\,\frac{2}{3}\left[C_A^3 (4-3 \zeta_3)+d_{44}
   (12+72 \zeta_3)\right]\frac{1}{\eps}
\, .
\end{align}
The above contribution is multiplied by a factor of $12$ while calculating $ \beta_{ffs}(a)$,  which cancels exactly the error term for \neqt SYM in tab.~\ref{tab:wrongres} at $\mathcal{O}(a^4)$, thereby restoring SUSY.

\begin{figure}[t!]
\includegraphics[width=0.3\linewidth]{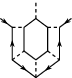}
\caption{\label{urdhpund} 
Diagram where Levi‑Civita contributions appear for the first time in \neqf SYM, resembling the \emph{Urdhwa Pundra} (upward lotus) symbol of Lord Vishnu in Hindu mythology.
}
\end{figure}
\paragraph*{\textbf{Implementation}} 
For the full computation of the theories, we generate the diagrams up to three loops using \texttt{Qgraf}~\cite{Nogueira:1991ex}. 
The symbolic manipulations are performed with the computer algebra system \texttt{FORM}~\cite{Vermaseren:2000nd,Kuipers:2012rf,Ruijl:2017dtg}, which we use to implement the Feynman rules, identify topologies, and carry out the colour, flavour, and spinor traces, this time including the correct Levi-Civita tensors. 
The loop integrations are computed with \texttt{FORCER}~\cite{Ruijl:2017cxj}, which is capable of evaluating two‑point functions up to four loops. 
The renormalisation of the bare quantities in the $\overline{\rm MS}$-scheme proceeds in the standard way.
As a check of the entire setup, we have also performed the renormalisation of the \neqqcd SUSY case, namely Quantum Chromodynamics (QCD), in standard DREG up to four loops~\cite{vanRitbergen:1997va,Czakon:2004bu} (for QCD $\beta$ function computations in DR, see~\cite{Jack:1994bn,Harlander:2006xq}).
Throughout the calculation a general covariant gauge was used, and we verified explicitly that all final results of the $\beta$-function are independent of the gauge parameter. 
Finally, substituting the appropriate values of $(n_s,n_f)$ yields SUSY‑consistent results for all \neqa  SYM theories. 

It appears that \neqo SYM, which lacks any flavour Levi‑Civita tensors, should be free of the $\gamma_5$ problem at all loop orders, see also~\cite{Stockinger:2005gx,Belusca-Maito:2023wah}.
In contrast, \neqf SYM could in principle be affected when six flavour matrices appear, which can happen only in the five‑loop topologies such as the diagram shown in fig~\ref{urdhpund}. 
To verify this, we have evaluated the four-loop $\beta$-functions for \neqsafe SYM,  dropping the Levi-Civita terms. 
We find 
\begin{align}
\label{eq:4loop-beta}
    \beta_{\mathcal{N}=1} \;&= -\left(6 a^2 C_A+12 a^3 C_A^2+42 a^4 C_A^3 + 204a^5C_A^4\right)\;,\nonumber \\
    \beta_{\mathcal{N}=4} \;&= 0\;.
\end{align}
Here, both the Yukawa and gauge $\beta$ functions are equal, i.e., eq.~\eqref{eq:betas} holds. 
The four-loop \neqo gauge $\beta_{ccg}$ was calculated in \cite{Harlander:2006xq} and agrees with eq.\eqref{eq:4loop-beta}, whereas the four-loop calculation of the \neqo Yukawa $\beta_{ffs}$ function in DR performed by us is a novel result.
The \neqf case matches with \cite{Velizhanin:2010vw}.
The four-loop \neqt case needs a non-trivial handling of the $\gamma_5$ problem and will be the topic of a future work.

\section{Conclusions}\label{sec:conc}
\if{1==0}
We have shown that the apparent failure of DREG in supersymmetric theories was caused by subtleties in the treatment of Levi‑Civita tensors. 
By introducing a generalized supersymmetric version of DREG, valid beyond the usual DR framework, we demonstrated up to three loops that SUSY‑consistent results are obtained, provided the $\gamma_5$ regularisation is handled consistently. 
This also allowed us to resolve the paradox concerning the $\beta$ function from the Yukawa interaction, explaining why only \neqt SYM was affected by the earlier computations in the literature, while results for \neqsafe SYM remained consistent.

Immediate generalizations include extending the renormalisation of $\mathcal{N}=1, 2$ and $4$ SYM theories to four and five loops. 
\fi

We have shown that the alleged failure of DREG in SYM theories was caused by subtleties in the treatment of Levi‑Civita tensors, correcting which restored SUSY at three loops. 
By using the formalism of the generalised Lagrangian, we were able to quantify the source of the apparent breakdown of supersymmetric regularisation.
This helped us go beyond just fixing the three-loop result, and enabled us to make general higher loop predictions for the regularisation of the \neqa SYM theories. We then carried out four-loop calculations that verified our predictions, in addition to producing a novel result.

We have demonstrated here that, there is no inherent problem in supersymmetric regularisation, the only problem is regularising $\gamma_5$ which is not a SUSY specific problem.
Thus, the generalized SUSY DREG framework suggests intriguing possibilities for formulating supersymmetric theories directly in $D = 4 - 2\eps$ dimensions, with potential implications for studies in the planar (’t Hooft) limit, where simplifications of colour structures often allow for exact or quasi‑exact results. 
It may also prove relevant for applying DREG to other highly symmetric theories with extended scalar sectors and Yukawa interactions, such as half-maximal ($\mathcal{N}=4$) or maximal ($\mathcal{N}=8$) supergravity and questions related to their ultraviolet finiteness, which involve double copies of the \neqf SYM theory.

\begin{acknowledgments}
We would like to thank John Gracey and Vitaly Velizhanin for useful discussions and S.M. acknowledges earlier collaboration with Vsevolod Chestnov.
The work has been supported by the European Research Council 
through  ERC Advanced Grant 101095857, {\it Conformal-EIC}.
\end{acknowledgments}

\newpage
\bibliographystyle{apsrev4-2}

\begin{thebibliography}{32}%
\makeatletter
\providecommand \@ifxundefined [1]{%
 \@ifx{#1\undefined}
}%
\providecommand \@ifnum [1]{%
 \ifnum #1\expandafter \@firstoftwo
 \else \expandafter \@secondoftwo
 \fi
}%
\providecommand \@ifx [1]{%
 \ifx #1\expandafter \@firstoftwo
 \else \expandafter \@secondoftwo
 \fi
}%
\providecommand \natexlab [1]{#1}%
\providecommand \enquote  [1]{``#1''}%
\providecommand \bibnamefont  [1]{#1}%
\providecommand \bibfnamefont [1]{#1}%
\providecommand \citenamefont [1]{#1}%
\providecommand \href@noop [0]{\@secondoftwo}%
\providecommand \href [0]{\begingroup \@sanitize@url \@href}%
\providecommand \@href[1]{\@@startlink{#1}\@@href}%
\providecommand \@@href[1]{\endgroup#1\@@endlink}%
\providecommand \@sanitize@url [0]{\catcode `\\12\catcode `\$12\catcode
  `\&12\catcode `\#12\catcode `\^12\catcode `\_12\catcode `\%12\relax}%
\providecommand \@@startlink[1]{}%
\providecommand \@@endlink[0]{}%
\providecommand \url  [0]{\begingroup\@sanitize@url \@url }%
\providecommand \@url [1]{\endgroup\@href {#1}{\urlprefix }}%
\providecommand \urlprefix  [0]{URL }%
\providecommand \Eprint [0]{\href }%
\providecommand \doibase [0]{https://doi.org/}%
\providecommand \selectlanguage [0]{\@gobble}%
\providecommand \bibinfo  [0]{\@secondoftwo}%
\providecommand \bibfield  [0]{\@secondoftwo}%
\providecommand \translation [1]{[#1]}%
\providecommand \BibitemOpen [0]{}%
\providecommand \bibitemStop [0]{}%
\providecommand \bibitemNoStop [0]{.\EOS\space}%
\providecommand \EOS [0]{\spacefactor3000\relax}%
\providecommand \BibitemShut  [1]{\csname bibitem#1\endcsname}%
\let\auto@bib@innerbib\@empty
%</preamble>
\bibitem [{\citenamefont {Martin}(1998)}]{Martin:1997ns}%
  \BibitemOpen
  \bibfield  {author} {\bibinfo {author} {\bibfnamefont {S.~P.}\ \bibnamefont
  {Martin}},\ }\href {https://doi.org/10.1142/9789812839657_0001} {\bibfield
  {journal} {\bibinfo  {journal} {Adv. Ser. Direct. High Energy Phys.}\
  }\textbf {\bibinfo {volume} {18}},\ \bibinfo {pages} {1} (\bibinfo {year}
  {1998})},\ \Eprint {https://arxiv.org/abs/hep-ph/9709356}
  {arXiv:hep-ph/9709356} \BibitemShut {NoStop}%
\bibitem [{\citenamefont {Novikov}\ \emph {et~al.}(1983)\citenamefont
  {Novikov}, \citenamefont {Shifman}, \citenamefont {Vainshtein},\ and\
  \citenamefont {Zakharov}}]{Novikov:1983uc}%
  \BibitemOpen
  \bibfield  {author} {\bibinfo {author} {\bibfnamefont {V.~A.}\ \bibnamefont
  {Novikov}}, \bibinfo {author} {\bibfnamefont {M.~A.}\ \bibnamefont
  {Shifman}}, \bibinfo {author} {\bibfnamefont {A.~I.}\ \bibnamefont
  {Vainshtein}},\ and\ \bibinfo {author} {\bibfnamefont {V.~I.}\ \bibnamefont
  {Zakharov}},\ }\href {https://doi.org/10.1016/0550-3213(83)90338-3}
  {\bibfield  {journal} {\bibinfo  {journal} {Nucl. Phys. B}\ }\textbf
  {\bibinfo {volume} {229}},\ \bibinfo {pages} {381} (\bibinfo {year}
  {1983})}\BibitemShut {NoStop}%
\bibitem [{\citenamefont {Avdeev}\ \emph {et~al.}(1980)\citenamefont {Avdeev},
  \citenamefont {Tarasov},\ and\ \citenamefont {Vladimirov}}]{Avdeev:1980bh}%
  \BibitemOpen
  \bibfield  {author} {\bibinfo {author} {\bibfnamefont {L.~V.}\ \bibnamefont
  {Avdeev}}, \bibinfo {author} {\bibfnamefont {O.~V.}\ \bibnamefont
  {Tarasov}},\ and\ \bibinfo {author} {\bibfnamefont {A.~A.}\ \bibnamefont
  {Vladimirov}},\ }\href {https://doi.org/10.1016/0370-2693(80)90219-1}
  {\bibfield  {journal} {\bibinfo  {journal} {Phys. Lett. B}\ }\textbf
  {\bibinfo {volume} {96}},\ \bibinfo {pages} {94} (\bibinfo {year}
  {1980})}\BibitemShut {NoStop}%
\bibitem [{\citenamefont {Grisaru}\ \emph {et~al.}(1980)\citenamefont
  {Grisaru}, \citenamefont {Rocek},\ and\ \citenamefont
  {Siegel}}]{Grisaru:1980nk}%
  \BibitemOpen
  \bibfield  {author} {\bibinfo {author} {\bibfnamefont {M.~T.}\ \bibnamefont
  {Grisaru}}, \bibinfo {author} {\bibfnamefont {M.}~\bibnamefont {Rocek}},\
  and\ \bibinfo {author} {\bibfnamefont {W.}~\bibnamefont {Siegel}},\ }\href
  {https://doi.org/10.1103/PhysRevLett.45.1063} {\bibfield  {journal} {\bibinfo
   {journal} {Phys. Rev. Lett.}\ }\textbf {\bibinfo {volume} {45}},\ \bibinfo
  {pages} {1063} (\bibinfo {year} {1980})}\BibitemShut {NoStop}%
\bibitem [{\citenamefont {'t~Hooft}\ and\ \citenamefont
  {Veltman}(1972)}]{tHooft:1972tcz}%
  \BibitemOpen
  \bibfield  {author} {\bibinfo {author} {\bibfnamefont {G.}~\bibnamefont
  {'t~Hooft}}\ and\ \bibinfo {author} {\bibfnamefont {M.~J.~G.}\ \bibnamefont
  {Veltman}},\ }\href {https://doi.org/10.1016/0550-3213(72)90279-9} {\bibfield
   {journal} {\bibinfo  {journal} {Nucl. Phys. B}\ }\textbf {\bibinfo {volume}
  {44}},\ \bibinfo {pages} {189} (\bibinfo {year} {1972})}\BibitemShut
  {NoStop}%
\bibitem [{\citenamefont {Bollini}\ and\ \citenamefont
  {Giambiagi}(1972)}]{Bollini:1972ui}%
  \BibitemOpen
  \bibfield  {author} {\bibinfo {author} {\bibfnamefont {C.~G.}\ \bibnamefont
  {Bollini}}\ and\ \bibinfo {author} {\bibfnamefont {J.~J.}\ \bibnamefont
  {Giambiagi}},\ }\href {https://doi.org/10.1007/BF02895558} {\bibfield
  {journal} {\bibinfo  {journal} {Nuovo Cim. B}\ }\textbf {\bibinfo {volume}
  {12}},\ \bibinfo {pages} {20} (\bibinfo {year} {1972})}\BibitemShut {NoStop}%
\bibitem [{\citenamefont {Siegel}(1979)}]{Siegel:1979wq}%
  \BibitemOpen
  \bibfield  {author} {\bibinfo {author} {\bibfnamefont {W.}~\bibnamefont
  {Siegel}},\ }\href {https://doi.org/10.1016/0370-2693(79)90282-X} {\bibfield
  {journal} {\bibinfo  {journal} {Phys. Lett. B}\ }\textbf {\bibinfo {volume}
  {84}},\ \bibinfo {pages} {193} (\bibinfo {year} {1979})}\BibitemShut
  {NoStop}%
\bibitem [{\citenamefont {Siegel}(1980)}]{Siegel:1980qs}%
  \BibitemOpen
  \bibfield  {author} {\bibinfo {author} {\bibfnamefont {W.}~\bibnamefont
  {Siegel}},\ }\href {https://doi.org/10.1016/0370-2693(80)90819-9} {\bibfield
  {journal} {\bibinfo  {journal} {Phys. Lett. B}\ }\textbf {\bibinfo {volume}
  {94}},\ \bibinfo {pages} {37} (\bibinfo {year} {1980})}\BibitemShut {NoStop}%
\bibitem [{\citenamefont {Avdeev}\ and\ \citenamefont
  {Tarasov}(1982)}]{Avdeev:1981ew}%
  \BibitemOpen
  \bibfield  {author} {\bibinfo {author} {\bibfnamefont {L.~V.}\ \bibnamefont
  {Avdeev}}\ and\ \bibinfo {author} {\bibfnamefont {O.~V.}\ \bibnamefont
  {Tarasov}},\ }\href {https://doi.org/10.1016/0370-2693(82)91068-1} {\bibfield
   {journal} {\bibinfo  {journal} {Phys. Lett. B}\ }\textbf {\bibinfo {volume}
  {112}},\ \bibinfo {pages} {356} (\bibinfo {year} {1982})}\BibitemShut
  {NoStop}%
\bibitem [{\citenamefont {Avdeev}(1982)}]{Avdeev:1982np}%
  \BibitemOpen
  \bibfield  {author} {\bibinfo {author} {\bibfnamefont {L.~V.}\ \bibnamefont
  {Avdeev}},\ }\href {https://doi.org/10.1016/0370-2693(82)90726-2} {\bibfield
  {journal} {\bibinfo  {journal} {Phys. Lett. B}\ }\textbf {\bibinfo {volume}
  {117}},\ \bibinfo {pages} {317} (\bibinfo {year} {1982})}\BibitemShut
  {NoStop}%
\bibitem [{\citenamefont {Velizhanin}(2009)}]{Velizhanin:2008rw}%
  \BibitemOpen
  \bibfield  {author} {\bibinfo {author} {\bibfnamefont {V.~N.}\ \bibnamefont
  {Velizhanin}},\ }\href {https://doi.org/10.1016/j.nuclphysb.2009.03.017}
  {\bibfield  {journal} {\bibinfo  {journal} {Nucl. Phys. B}\ }\textbf
  {\bibinfo {volume} {818}},\ \bibinfo {pages} {95} (\bibinfo {year} {2009})},\
  \Eprint {https://arxiv.org/abs/0809.2509} {arXiv:0809.2509 [hep-th]}
  \BibitemShut {NoStop}%
\bibitem [{\citenamefont {Chestnov}(2019)}]{Chestnov:2019bed}%
  \BibitemOpen
  \bibfield  {author} {\bibinfo {author} {\bibfnamefont {V.}~\bibnamefont
  {Chestnov}},\ }\emph {\bibinfo {title} {{High-precision computation of
  observables in supersymmetric Yang{\textendash}Mills theory}}},\ \href@noop
  {} {Ph.D. thesis},\ \bibinfo  {school} {Hamburg U.} (\bibinfo {year}
  {2019}),\ \bibinfo {note}
  {{\url{https://bib-pubdb1.desy.de/record/424578}}}\BibitemShut {NoStop}%
\bibitem [{\citenamefont {Velizhanin}(2011)}]{Velizhanin:2010vw}%
  \BibitemOpen
  \bibfield  {author} {\bibinfo {author} {\bibfnamefont {V.~N.}\ \bibnamefont
  {Velizhanin}},\ }\href {https://doi.org/10.1016/j.physletb.2011.01.019}
  {\bibfield  {journal} {\bibinfo  {journal} {Phys. Lett. B}\ }\textbf
  {\bibinfo {volume} {696}},\ \bibinfo {pages} {560} (\bibinfo {year}
  {2011})},\ \Eprint {https://arxiv.org/abs/1008.2198} {arXiv:1008.2198
  [hep-th]} \BibitemShut {NoStop}%
\bibitem [{\citenamefont {van Ritbergen}\ \emph {et~al.}(1999)\citenamefont
  {van Ritbergen}, \citenamefont {Schellekens},\ and\ \citenamefont
  {Vermaseren}}]{vanRitbergen:1998pn}%
  \BibitemOpen
  \bibfield  {author} {\bibinfo {author} {\bibfnamefont {T.}~\bibnamefont {van
  Ritbergen}}, \bibinfo {author} {\bibfnamefont {A.~N.}\ \bibnamefont
  {Schellekens}},\ and\ \bibinfo {author} {\bibfnamefont {J.~A.~M.}\
  \bibnamefont {Vermaseren}},\ }\href
  {https://doi.org/10.1142/S0217751X99000038} {\bibfield  {journal} {\bibinfo
  {journal} {Int. J. Mod. Phys. A}\ }\textbf {\bibinfo {volume} {14}},\
  \bibinfo {pages} {41} (\bibinfo {year} {1999})},\ \Eprint
  {https://arxiv.org/abs/hep-ph/9802376} {arXiv:hep-ph/9802376} \BibitemShut
  {NoStop}%
\bibitem [{\citenamefont {Chakraborty}\ and\ \citenamefont
  {Moch}(2026)}]{emgsusy}%
  \BibitemOpen
  \bibfield  {author} {\bibinfo {author} {\bibfnamefont {M.}~\bibnamefont
  {Chakraborty}}\ and\ \bibinfo {author} {\bibfnamefont {S.}~\bibnamefont
  {Moch}},\ }\href@noop {} {\bibinfo {title} {{\it{Connecting Supersymmetry to
  Non-Supersymmetric theories: The Gross-Neveu-Yukawa example}}}} (\bibinfo
  {year} {2026}),\ \bibinfo {note}
  {{\href{https://doi.org/10.3204/PUBDB-2026-00812}{PUBDB-2026-00812}}}\BibitemShut
  {NoStop}%
\bibitem [{\citenamefont {Davies}\ \emph {et~al.}(2020)\citenamefont {Davies},
  \citenamefont {Herren}, \citenamefont {Poole}, \citenamefont {Steinhauser},\
  and\ \citenamefont {Thomsen}}]{Davies:2019onf}%
  \BibitemOpen
  \bibfield  {author} {\bibinfo {author} {\bibfnamefont {J.}~\bibnamefont
  {Davies}}, \bibinfo {author} {\bibfnamefont {F.}~\bibnamefont {Herren}},
  \bibinfo {author} {\bibfnamefont {C.}~\bibnamefont {Poole}}, \bibinfo
  {author} {\bibfnamefont {M.}~\bibnamefont {Steinhauser}},\ and\ \bibinfo
  {author} {\bibfnamefont {A.~E.}\ \bibnamefont {Thomsen}},\ }\href
  {https://doi.org/10.1103/PhysRevLett.124.071803} {\bibfield  {journal}
  {\bibinfo  {journal} {Phys. Rev. Lett.}\ }\textbf {\bibinfo {volume} {124}},\
  \bibinfo {pages} {071803} (\bibinfo {year} {2020})},\ \Eprint
  {https://arxiv.org/abs/1912.07624} {arXiv:1912.07624 [hep-ph]} \BibitemShut
  {NoStop}%
\bibitem [{\citenamefont {Poole}\ and\ \citenamefont
  {Thomsen}(2019{\natexlab{a}})}]{Poole:2019txl}%
  \BibitemOpen
  \bibfield  {author} {\bibinfo {author} {\bibfnamefont {C.}~\bibnamefont
  {Poole}}\ and\ \bibinfo {author} {\bibfnamefont {A.~E.}\ \bibnamefont
  {Thomsen}},\ }\href {https://doi.org/10.1103/PhysRevLett.123.041602}
  {\bibfield  {journal} {\bibinfo  {journal} {Phys. Rev. Lett.}\ }\textbf
  {\bibinfo {volume} {123}},\ \bibinfo {pages} {041602} (\bibinfo {year}
  {2019}{\natexlab{a}})},\ \Eprint {https://arxiv.org/abs/1901.02749}
  {arXiv:1901.02749 [hep-th]} \BibitemShut {NoStop}%
\bibitem [{\citenamefont {Poole}\ and\ \citenamefont
  {Thomsen}(2019{\natexlab{b}})}]{Poole:2019kcm}%
  \BibitemOpen
  \bibfield  {author} {\bibinfo {author} {\bibfnamefont {C.}~\bibnamefont
  {Poole}}\ and\ \bibinfo {author} {\bibfnamefont {A.~E.}\ \bibnamefont
  {Thomsen}},\ }\href {https://doi.org/10.1007/JHEP09(2019)055} {\bibfield
  {journal} {\bibinfo  {journal} {JHEP}\ }\textbf {\bibinfo {volume} {09}},\
  \bibinfo {pages} {055}},\ \Eprint {https://arxiv.org/abs/1906.04625}
  {arXiv:1906.04625 [hep-th]} \BibitemShut {NoStop}%
\bibitem [{\citenamefont {Avdeev}\ and\ \citenamefont
  {Vladimirov}(1983)}]{Avdeev:1982xy}%
  \BibitemOpen
  \bibfield  {author} {\bibinfo {author} {\bibfnamefont {L.~V.}\ \bibnamefont
  {Avdeev}}\ and\ \bibinfo {author} {\bibfnamefont {A.~A.}\ \bibnamefont
  {Vladimirov}},\ }\href {https://doi.org/10.1016/0550-3213(83)90437-6}
  {\bibfield  {journal} {\bibinfo  {journal} {Nucl. Phys. B}\ }\textbf
  {\bibinfo {volume} {219}},\ \bibinfo {pages} {262} (\bibinfo {year}
  {1983})}\BibitemShut {NoStop}%
\bibitem [{\citenamefont {Vladimirov}(1980)}]{Vladimirov:1979zm}%
  \BibitemOpen
  \bibfield  {author} {\bibinfo {author} {\bibfnamefont {A.~A.}\ \bibnamefont
  {Vladimirov}},\ }\href {https://doi.org/10.1007/BF01018394} {\bibfield
  {journal} {\bibinfo  {journal} {Theor. Math. Phys.}\ }\textbf {\bibinfo
  {volume} {43}},\ \bibinfo {pages} {417} (\bibinfo {year} {1980})}\BibitemShut
  {NoStop}%
\bibitem [{\citenamefont {Chetyrkin}\ and\ \citenamefont
  {Tkachov}(1982)}]{Chetyrkin:1982nn}%
  \BibitemOpen
  \bibfield  {author} {\bibinfo {author} {\bibfnamefont {K.~G.}\ \bibnamefont
  {Chetyrkin}}\ and\ \bibinfo {author} {\bibfnamefont {F.~V.}\ \bibnamefont
  {Tkachov}},\ }\href {https://doi.org/10.1016/0370-2693(82)90358-6} {\bibfield
   {journal} {\bibinfo  {journal} {Phys. Lett. B}\ }\textbf {\bibinfo {volume}
  {114}},\ \bibinfo {pages} {340} (\bibinfo {year} {1982})}\BibitemShut
  {NoStop}%
\bibitem [{\citenamefont {Nogueira}(1993)}]{Nogueira:1991ex}%
  \BibitemOpen
  \bibfield  {author} {\bibinfo {author} {\bibfnamefont {P.}~\bibnamefont
  {Nogueira}},\ }\href {https://doi.org/10.1006/jcph.1993.1074} {\bibfield
  {journal} {\bibinfo  {journal} {J. Comput. Phys.}\ }\textbf {\bibinfo
  {volume} {105}},\ \bibinfo {pages} {279} (\bibinfo {year}
  {1993})}\BibitemShut {NoStop}%
\bibitem [{\citenamefont {Vermaseren}(2000)}]{Vermaseren:2000nd}%
  \BibitemOpen
  \bibfield  {author} {\bibinfo {author} {\bibfnamefont {J.~A.~M.}\
  \bibnamefont {Vermaseren}},\ }\href@noop {} {\bibinfo {title} {{New features
  of FORM}}} (\bibinfo {year} {2000}),\ \Eprint
  {https://arxiv.org/abs/math-ph/0010025} {arXiv:math-ph/0010025} \BibitemShut
  {NoStop}%
\bibitem [{\citenamefont {Kuipers}\ \emph {et~al.}(2013)\citenamefont
  {Kuipers}, \citenamefont {Ueda}, \citenamefont {Vermaseren},\ and\
  \citenamefont {Vollinga}}]{Kuipers:2012rf}%
  \BibitemOpen
  \bibfield  {author} {\bibinfo {author} {\bibfnamefont {J.}~\bibnamefont
  {Kuipers}}, \bibinfo {author} {\bibfnamefont {T.}~\bibnamefont {Ueda}},
  \bibinfo {author} {\bibfnamefont {J.~A.~M.}\ \bibnamefont {Vermaseren}},\
  and\ \bibinfo {author} {\bibfnamefont {J.}~\bibnamefont {Vollinga}},\ }\href
  {https://doi.org/10.1016/j.cpc.2012.12.028} {\bibfield  {journal} {\bibinfo
  {journal} {Comput. Phys. Commun.}\ }\textbf {\bibinfo {volume} {184}},\
  \bibinfo {pages} {1453} (\bibinfo {year} {2013})},\ \Eprint
  {https://arxiv.org/abs/1203.6543} {arXiv:1203.6543 [cs.SC]} \BibitemShut
  {NoStop}%
\bibitem [{\citenamefont {Ruijl}\ \emph {et~al.}(2017)\citenamefont {Ruijl},
  \citenamefont {Ueda},\ and\ \citenamefont {Vermaseren}}]{Ruijl:2017dtg}%
  \BibitemOpen
  \bibfield  {author} {\bibinfo {author} {\bibfnamefont {B.}~\bibnamefont
  {Ruijl}}, \bibinfo {author} {\bibfnamefont {T.}~\bibnamefont {Ueda}},\ and\
  \bibinfo {author} {\bibfnamefont {J.}~\bibnamefont {Vermaseren}},\
  }\href@noop {} {\bibinfo {title} {{\texttt{FORM} version 4.2}}} (\bibinfo
  {year} {2017}),\ \Eprint {https://arxiv.org/abs/1707.06453} {arXiv:1707.06453
  [hep-ph]} \BibitemShut {NoStop}%
\bibitem [{\citenamefont {Ruijl}\ \emph {et~al.}(2020)\citenamefont {Ruijl},
  \citenamefont {Ueda},\ and\ \citenamefont {Vermaseren}}]{Ruijl:2017cxj}%
  \BibitemOpen
  \bibfield  {author} {\bibinfo {author} {\bibfnamefont {B.}~\bibnamefont
  {Ruijl}}, \bibinfo {author} {\bibfnamefont {T.}~\bibnamefont {Ueda}},\ and\
  \bibinfo {author} {\bibfnamefont {J.~A.~M.}\ \bibnamefont {Vermaseren}},\
  }\href {https://doi.org/10.1016/j.cpc.2020.107198} {\bibfield  {journal}
  {\bibinfo  {journal} {Comput. Phys. Commun.}\ }\textbf {\bibinfo {volume}
  {253}},\ \bibinfo {pages} {107198} (\bibinfo {year} {2020})},\ \Eprint
  {https://arxiv.org/abs/1704.06650} {arXiv:1704.06650 [hep-ph]} \BibitemShut
  {NoStop}%
\bibitem [{\citenamefont {van Ritbergen}\ \emph {et~al.}(1997)\citenamefont
  {van Ritbergen}, \citenamefont {Vermaseren},\ and\ \citenamefont
  {Larin}}]{vanRitbergen:1997va}%
  \BibitemOpen
  \bibfield  {author} {\bibinfo {author} {\bibfnamefont {T.}~\bibnamefont {van
  Ritbergen}}, \bibinfo {author} {\bibfnamefont {J.~A.~M.}\ \bibnamefont
  {Vermaseren}},\ and\ \bibinfo {author} {\bibfnamefont {S.~A.}\ \bibnamefont
  {Larin}},\ }\href {https://doi.org/10.1016/S0370-2693(97)00370-5} {\bibfield
  {journal} {\bibinfo  {journal} {Phys. Lett. B}\ }\textbf {\bibinfo {volume}
  {400}},\ \bibinfo {pages} {379} (\bibinfo {year} {1997})},\ \Eprint
  {https://arxiv.org/abs/hep-ph/9701390} {arXiv:hep-ph/9701390} \BibitemShut
  {NoStop}%
\bibitem [{\citenamefont {Czakon}(2005)}]{Czakon:2004bu}%
  \BibitemOpen
  \bibfield  {author} {\bibinfo {author} {\bibfnamefont {M.}~\bibnamefont
  {Czakon}},\ }\href {https://doi.org/10.1016/j.nuclphysb.2005.01.012}
  {\bibfield  {journal} {\bibinfo  {journal} {Nucl. Phys. B}\ }\textbf
  {\bibinfo {volume} {710}},\ \bibinfo {pages} {485} (\bibinfo {year}
  {2005})},\ \Eprint {https://arxiv.org/abs/hep-ph/0411261}
  {arXiv:hep-ph/0411261} \BibitemShut {NoStop}%
\bibitem [{\citenamefont {Jack}\ \emph {et~al.}(1994)\citenamefont {Jack},
  \citenamefont {Jones},\ and\ \citenamefont {Roberts}}]{Jack:1994bn}%
  \BibitemOpen
  \bibfield  {author} {\bibinfo {author} {\bibfnamefont {I.}~\bibnamefont
  {Jack}}, \bibinfo {author} {\bibfnamefont {D.~R.~T.}\ \bibnamefont {Jones}},\
  and\ \bibinfo {author} {\bibfnamefont {K.~L.}\ \bibnamefont {Roberts}},\
  }\href {https://doi.org/10.1007/BF01577555} {\bibfield  {journal} {\bibinfo
  {journal} {Z. Phys. C}\ }\textbf {\bibinfo {volume} {63}},\ \bibinfo {pages}
  {151} (\bibinfo {year} {1994})},\ \Eprint
  {https://arxiv.org/abs/hep-ph/9401349} {arXiv:hep-ph/9401349} \BibitemShut
  {NoStop}%
\bibitem [{\citenamefont {Harlander}\ \emph {et~al.}(2006)\citenamefont
  {Harlander}, \citenamefont {Jones}, \citenamefont {Kant}, \citenamefont
  {Mihaila},\ and\ \citenamefont {Steinhauser}}]{Harlander:2006xq}%
  \BibitemOpen
  \bibfield  {author} {\bibinfo {author} {\bibfnamefont {R.~V.}\ \bibnamefont
  {Harlander}}, \bibinfo {author} {\bibfnamefont {D.~R.~T.}\ \bibnamefont
  {Jones}}, \bibinfo {author} {\bibfnamefont {P.}~\bibnamefont {Kant}},
  \bibinfo {author} {\bibfnamefont {L.}~\bibnamefont {Mihaila}},\ and\ \bibinfo
  {author} {\bibfnamefont {M.}~\bibnamefont {Steinhauser}},\ }\href
  {https://doi.org/10.1088/1126-6708/2006/12/024} {\bibfield  {journal}
  {\bibinfo  {journal} {JHEP}\ }\textbf {\bibinfo {volume} {12}},\ \bibinfo
  {pages} {024}},\ \Eprint {https://arxiv.org/abs/hep-ph/0610206}
  {arXiv:hep-ph/0610206} \BibitemShut {NoStop}%
\bibitem [{\citenamefont {St{\"o}ckinger}(2005)}]{Stockinger:2005gx}%
  \BibitemOpen
  \bibfield  {author} {\bibinfo {author} {\bibfnamefont {D.}~\bibnamefont
  {St{\"o}ckinger}},\ }\href {https://doi.org/10.1088/1126-6708/2005/03/076}
  {\bibfield  {journal} {\bibinfo  {journal} {JHEP}\ }\textbf {\bibinfo
  {volume} {03}},\ \bibinfo {pages} {076}},\ \Eprint
  {https://arxiv.org/abs/hep-ph/0503129} {arXiv:hep-ph/0503129} \BibitemShut
  {NoStop}%
\bibitem [{\citenamefont {B{\'e}lusca-Ma{\"\i}to}\ \emph
  {et~al.}(2023)\citenamefont {B{\'e}lusca-Ma{\"\i}to}, \citenamefont
  {Ilakovac}, \citenamefont {K{\"u}hler}, \citenamefont
  {Mador-Bo{\v{z}}inovi{\'c}}, \citenamefont {St{\"o}ckinger},\ and\
  \citenamefont {Wei{\ss}wange}}]{Belusca-Maito:2023wah}%
  \BibitemOpen
  \bibfield  {author} {\bibinfo {author} {\bibfnamefont {H.}~\bibnamefont
  {B{\'e}lusca-Ma{\"\i}to}}, \bibinfo {author} {\bibfnamefont {A.}~\bibnamefont
  {Ilakovac}}, \bibinfo {author} {\bibfnamefont {P.}~\bibnamefont
  {K{\"u}hler}}, \bibinfo {author} {\bibfnamefont {M.}~\bibnamefont
  {Mador-Bo{\v{z}}inovi{\'c}}}, \bibinfo {author} {\bibfnamefont
  {D.}~\bibnamefont {St{\"o}ckinger}},\ and\ \bibinfo {author} {\bibfnamefont
  {M.}~\bibnamefont {Wei{\ss}wange}},\ }\href
  {https://doi.org/10.3390/sym15030622} {\bibfield  {journal} {\bibinfo
  {journal} {Symmetry}\ }\textbf {\bibinfo {volume} {15}},\ \bibinfo {pages}
  {622} (\bibinfo {year} {2023})},\ \Eprint {https://arxiv.org/abs/2303.09120}
  {arXiv:2303.09120 [hep-ph]} \BibitemShut {NoStop}%
\end{thebibliography}

\end{document}